\documentclass[epj]{svjour}
\usepackage{graphics}
\usepackage{amsmath}
\usepackage{graphicx}
\usepackage{subfigure}
\usepackage{amsfonts}
\usepackage{color}

\begin{document}

\title{Lattice Boltzmann study of chemically-driven self-propelled droplets}

\author{F. Fadda\inst{a}, G. Gonnella\inst{a} \and A. Lamura\inst{b} \and A. Tiribocchi\inst{c}}           
\offprints{\\ \\fede.fadda1110@gmail.com,\\ giuseppe.gonnella@ba.infn.it,\\ a.lamura@ba.iac.cnr.it,\\ adriano.tiribocchi@pd.infn.it}         
\institute{\inst{a}Dipartimento di Fisica and Sezione INFN Bari, Via Amendola 173, 70126 Bari, Italy,\\ \inst{b}Istituto Applicazioni Calcolo, CNR, Via Amendola 122/D, 70126 Bari, Italy,\\ \inst{c}Dipartimento di Fisica e Astronomia Via Marzolo 8, I-35131 Padova, Italy}

\date{Received: date / Revised version: date}

\abstract{
We numerically study the behavior of self-propelled liquid droplets whose motion is triggered by a Marangoni-like flow. This latter is generated by
variations of surfactant concentration which affect the droplet surface tension promoting its motion. In the present
paper a model for droplets with a third amphiphilic component is adopted.
The dynamics is described by Navier-Stokes and convection-diffusion equations, solved by lattice Boltzmann method coupled with finite-difference schemes.
We focus on two cases. First the study of self-propulsion of an isolated droplet is carried on and, then, 
the interaction 
of two self-propelled droplets is investigated.
In both cases, when the surfactant migrates towards the interface, a quadrupolar vortex of the velocity field forms inside the droplet and causes 
the motion. A weaker dipolar field emerges instead  when the surfactant is mainly diluted in the bulk. 
The dynamics of two interacting droplets is more complex and strongly depends on their reciprocal distance.  
If, in a head-on collision, droplets are close enough, the velocity field initially 
attracts them
until a motionless steady state is achieved. If the droplets are vertically shifted, the hydrodynamic field leads to an initial reciprocal attraction followed by a 
scattering along
opposite directions. This hydrodynamic interaction acts on a separation of some droplet radii otherwise it becomes 
negligible and droplets
motion is only driven by Marangoni effect. Finally, if one of the droplets is passive, this latter is generally advected by the fluid flow generated by the active one.
} 
\maketitle

\section{Introduction}\label{intro}
Much research has been recently focussed on the study of self-propelled droplets, a remarkable example of system that shows autonomous dissipative motion sustained by an internal energy supply~\cite{1,2}.
Various mechanisms triggering droplet locomotion have been proposed during years. These rely, for instance, on the phase separation of a multicomponent fluid~\cite{3}, on redox reactions in an 
electrolyte medium~\cite{4}, or on chemical reactions such as the  Belousov-Zhabotinskii ones, in which traveling waves of chemical activity inside the droplet determine an oscillation 
of the interface which favours the self-propulsion~\cite{6,7}. Importantly the presence of spatial heterogeneities in the distribution of a chemically reactive solution is crucial to generate
spontaneous motion.

Autonomous motile droplets are of particular interest in biology as they can mimic, for example, the spontaneous motion of cells~\cite{8,9} or the behavior of protozoal organisms~\cite{10}.
The latter include organisms particularly abundant in nature such as cyanobacteria, {\it paramecium} and {\it volvox}, whose motion is guided by tangential or radial deformations of their
surface~\cite{ehlers,drescher}. In a recent work Thutupalli et al.~\cite{13} have shown that a spontaneous motion of self-propelled non-living droplets
can be driven by chemical reactions based on a symmetry-breaking mechanism. More specifically, the motion is triggered by 
a Marangoni flow~\cite{14,15,16} which results from the spontaneous bromination of a surfactant located on the interface of a water droplet dispersed into an oil background.
This, in turn, leads to an anisotropic surface tension which creates the flow inside and outside the droplet. 
The self-propulsion using Marangoni effect was initially proposed by Ryazantsev~\cite{19} and is now widely used to create self-motile objects~\cite{13,20,21,22,23,26,29}. 
Such mechanism has been theoretically modeled in several studies~\cite{37,38,40,41,42,43,44}, usually through a set of balance equations in which the evolution of the surfactant
is governed by a convection-diffusion equation (in which production and consumption contributions are included) 
while its advection towards (or away from) the droplet interface is driven by the velocity of the fluid.  The latter instead obeys the Stokes equation. The change in the concentration of the material is typically described by a further convection-diffusion equation.
While many theoretical studies have shed light either on the chemical reactions occurring in the system or on the symmetry breaking mechanisms promoting the droplet motion, 
much less is known about the structure of the hydrodynamic fluid flow, inside and in the surroundings of the droplet. The interplay between the local velocity field and the
concentration of surfactant is essential to understand how the motion occurs and, eventually, to suggest experimental procedures to control it. This is even more important for 
an emulsion of droplets, as it could provide insights for the design of complex structures of technological interest starting from a primordial disordered arrangement of droplets. 

The purpose of the present paper is to elucidate the role played by hydrodynamic interactions when droplets, in an extremely diluted regime, acquire motion through a Marangoni-like flow.
This is created by an inhomogeneous distribution of surfactant inside the droplet which, in turn, affects the surface tension.
We have found that the velocity field structure has a non-trivial dependence on the dynamics of the surfactant. 
If this is produced in a circular region inside the droplet (tangentially to the interface) and is then diluted in the bulk of the fluid (without being absorbed at the interface), a dipolar velocity field forms 
inside the droplet itself, which then moves unidirectionally, where the surface tension is higher. This flow is however much weaker than that observed when the surfactant migrates towards the interface. 
In this case an intense flow, distributed tangentially to the droplet interface (where the local surface tension is lower), pushes the local fluid backwards and favours the formation of four vortices located inside it. Interestingly, in all cases two vortices form outside the droplet in its front and favour a unidirectional motion.
We have also extended the study to the case of two interacting self-propelled droplets, which, to the best of our knowledge, has been theoretically investigated only recently for the first time~\cite{42}.
Several system realizations have been studied. For instance, if the droplets are collinear and move against each other,
a steady state is achieved due to a balance between the Marangoni flow,
promoting their collision, and the flow field formed in between the droplets, favouring repulsion. If the droplets are shifted vertically and are close enough, a mutual attraction is initially
found followed, later on, by a reciprocal repulsion due to the hydrodynamic field emerging between them. If they are vertically aligned and again sufficiently close, the repulsion initially dominates 
while at late times the motion is driven only by Marangoni effect as the interaction between droplets becomes negligible.
If only one of the two droplets is active, the passive droplet is usually advected by the fluid flow generated by the active one.

We finally mention that the thermodynamic description adopted here is the same as that presented in Ref.~\cite{41} for the single self-propelled droplet and in Ref.~\cite{42} for 
two colliding self-propelled droplets. However, in the analytical study of  Ref.~\cite{41} (i) hydrodynamic effects are taken into account 
in the Stokes approximation and (ii) the structure of the velocity field and its effect on the dynamics of both the concentration of the fluid and of the surfactant is not numerically discussed.
Also, in Ref.~\cite{42} only head-on collisions are considered, whereas here the study has been extended to further cases.
The advantage of our model is that it represents an intentionally simplified description of complex biochemical processes which allows us to isolate a few key physical parameters 
controlling the hydrodynamics in the system.

The paper is structured as follows. In Sec. 2 we describe the thermo-hydrodynamics of the system and the numerical method adopted to numerically solve the hydrodynamic equations governing its physics.
In Sec. 3 we discuss the equilibrium properties of a planar interface and of a motionless droplet when the surfactant is either absorbed at the interface or not. Sec. 4 is dedicated to present the
results of a single motile droplet whose propulsion is triggered by a surfactant produced in a circular region tangential to its interface. 
In a further section we report the results regarding the interaction of two self-propelled droplets, both when the reciprocal motion is collinear and when it is vertically shifted. 
A final section is devoted to the discussion of the results and to the conclusions.

\section{The model}
In this Section we first describe the equilibrium model of the fluid mixture with surfactant. We then introduce the phenomenological equations
governing the physics in which the contribution of source and consumption terms of the surfactant is added. Finally we give details about the adopted numerical model.

\subsection{Thermo-hydrodynamic description}
We first consider a passive ternary fluid mixture with equilibrium free energy \cite{41}
\begin{eqnarray}
\mathcal{F}&=&\int d\mathbf{r}~f(n,\phi,c)\nonumber\\
&=& \int d\mathbf{r}\left [ n T \ln n + f_{GL}(\phi) + \frac{B(c)}{2} (\nabla\phi)^{2} + c\ln c \right ],
\end{eqnarray}
where $f(n,\phi,c)$ is the total free energy density, $\phi$ is the local concentration difference of two immiscible components of the mixture, 
$c$ is the concentration of surfactant (the third component), $T$ is the temperature, assumed fixed in the following, 
and $n$ is the total density of the mixture. 
The term depending on $n$ gives rise to the ideal gas pressure $p^{i}=nT$ which does not affect the phase behavior.
The free-energy density $f_{GL}(\phi)=\frac{a}{2}\phi^{2}+\frac{b}{4}\phi^{4}$ corresponds to the polynomial part
of a Ginzburg-Landau free energy describing
the bulk properties of the system \cite{45}. The coefficient $b$ is always positive while $a$ allows  to discriminate between
a homogeneous ($a>0$) mixture  and a separated one ($a<0$), where the two components
coexist with equilibrium values $\phi=\pm\phi^{eq}$ being $\phi^{eq}=\sqrt{-a/b}$.
The interfacial properties of the mixture are controlled by the gradient term
where the coupling $B(c)$ is assumed globally
positive with expression  $B(c)=B_{0}+B_{1}c$ \cite{41}.  
The coefficient $B_{1}$ determines whether the surfactant migrates to the interface ($B_{1}<0$) or away from it in the bulk ($B_{1}>0$).  
The logarithmic term $f_{0}=c \ln c$ arises from the translational entropy of the dilute amphiphilic component. 

The time-evolution of $\phi$ is described by the convection-diffusion equation 
\begin{equation}
\label{Equ.Phi.}
\partial_{t}\phi+\nabla \cdot (\phi \mathbf{u})=\nabla^{2}\mu_{\phi}	
\end{equation}
where $\mu_{\phi}$ is the chemical potential given by
\begin{equation}
\mu_{\phi}
=\frac{\delta \mathcal{F}}{\delta \phi}=a\phi+b\phi^{3}-B(c)\nabla^{2}\phi-B_{1}\nabla\phi\cdot\nabla c .
\end{equation}
The dilute component $c$ obeys a convection-diffusion equation as well
\begin{equation}
\label{Equ.Surf.}
\partial_{t}c+\nabla \cdot (c \mathbf{u})=\nabla \cdot \left [ L(c)\nabla \mu_c \right ]
\end{equation}
where $L(c)=Dc$ and $D$ is the diffusion coefficient for the surfactant. 
The quantity $\mu_c$ is the chemical potential of the surfactant 
\begin{equation}\label{chem_c}
\mu_c=\frac{\delta\mathcal{F}}{\delta c}=\ln c+1+\frac{B_{1}}{2}(\nabla\phi)^{2} .
\end{equation}
In the right hand side of Eq.~(\ref{Equ.Surf.}) the linear dependence of $L(c)$ is necessary for a dilute component \cite{41},
in order to avoid a singularity at $c=0$ in the surfactant density current ${\bf j}_c=-L(c)\nabla\mu_c$ (if the r.h.s of Eq.~(\ref{Equ.Surf.}) is written as $-\nabla\cdot{\bf j}_c$).

The concentrations $\phi$ and $c$ are coupled to the local velocity
\textbf{u} of the whole fluid which obeys the Navier-Stokes equation
\begin{equation}
\begin{split}
\label{Navier-Stokes}
\partial_{t}(nu_{\beta})+\partial_{\alpha}(nu_{\alpha}u_{\beta})=-\partial_{\alpha}P_{\alpha\beta}\\
+\partial_{\alpha}\left \{ \eta\left ( \partial_{\alpha}u_{\beta}+\partial_{\beta}u_{\alpha}-2\frac{\delta_{\alpha\beta}}{d}\partial_{\gamma}u_{\gamma}\right )+\zeta\delta_{\alpha\beta} \partial_{\gamma}u_{\gamma}\right \}
\end{split}
\end{equation}
where $\eta$ and $\zeta$ are the shear and bulk viscosities respectively and $d$ is the dimensionality of the system ($d=2$ in our case).
The fluid mass density $n$ satisfies the continuity equation
\begin{equation}
\label{Continuity equation}
\partial_{t}n+\nabla \cdot (n \mathbf{u})=0 .
\end{equation}
$P_{\alpha\beta}$ is the pressure tensor of the system. Its expression is
obtained by
\begin{equation}
\begin{split}
P_{\alpha\beta}=\left [ \phi\frac{\delta \mathcal{F}}{\delta\phi}+c\frac{\delta \mathcal{F}}{\delta c}+n\frac{\delta \mathcal{F}}{\delta n}-f(n,\phi,c)\right ]\delta_{\alpha\beta}+D_{\alpha\beta}
\end{split}
\end{equation}
where $D_{\alpha\beta}$ has to ensure the general equilibrium condition $\partial_{\alpha}P_{\alpha\beta}=0$.
It turns out that $P_{\alpha\beta}=p_{0}\delta_{\alpha\beta}+B(c)\partial_{\alpha}\phi\partial_{\beta}\phi$
with
\begin{equation}
\begin{split}
 p_{0}=p^i+\frac{a}{2}\phi^{2}+\frac{3b}{4}\phi^{4} - B(c) \phi \nabla^{2} \phi\\ - B_{1} \phi \nabla c \cdot \nabla \phi+
 c-\frac{B_{0}}{2} \nabla \phi\cdot\nabla \phi.
\end{split}
\end{equation}

The active character of the droplet is described by adding a source term
$A\Theta(R-|\mathbf{r}-\mathbf{r}_{P}|)$ and a consumption one
$-\gamma(c-c_{0})$ to the Eq.~(\ref{Equ.Surf.}) of the surfactant concentration.
The source term with coefficient $A>0$ mimicks 
the production of surfactant in a circular region, centered 
in $\textbf{r}_{P}$ inside the droplet, of radius R, $\Theta$ being the 
Heaviside step function.
The second term promotes a reduction of $c$ 
with rate $\gamma>0$, $c_{0}$ being the minimum value of $c$ 
all over the system.
This is slightly different from the case discussed in Ref.~\cite{41}, where a spontaneous motion is achieved when the distribution of the surfactant, initially covering
the whole droplet, is slightly shifted from the droplet center of mass.

Equations (\ref{Equ.Phi.}), (\ref{Equ.Surf.}), (\ref{Navier-Stokes}) and (\ref{Continuity equation}) are solved 
numerically by using a hybrid method~\cite{46}, with a standard lattice Boltzmann approach for 
the continuity and the Navier-Stokes equations and a finite difference scheme for the convection-diffusion equations.
The method is described in the next sections. 

\subsection{Lattice Boltzmann numerical scheme}
We adopt a bidimensional square lattice ($D_{2}Q_{9}$ geometry) of size $L_x \times L_y$.
Horizontal (H) and vertical (V) links have length $\Delta x$ and diagonal (D) ones $\sqrt{2} \Delta x$. 
On each lattice site $\textbf{r}$ nine lattice velocities $\textbf{e}_{i}$ are defined ($i=0,...,8$). 
They have modulus $|\textbf{e}_{i}|=\Delta x/\Delta t_{LB}=e$ for H and V directions and $|\textbf{e}_{i}|=\sqrt2 e$ for D directions
(with $\Delta t_{LB}$ the time step), and $|\textbf{e}_{0}|=0$ for the rest velocity.
A set of distribution functions $f_{i}(\textbf{r},t)$ is defined on each lattice site at each time t. 
They evolve according to a single relaxation-time Boltzmann equation with a forcing term $F_{i}$
\begin{equation}
\begin{split}
f_{i}(\mathbf{r}+\mathbf{e}_{i}\Delta t_{LB},t+\Delta t_{LB})-f_{i}(\mathbf{r},t)=\\
-\frac{\Delta t_{LB}}{\tau}[f_{i}(\mathbf{r},t)-f_{i}^{eq}(\mathbf{r},t)]+\Delta t_{LB}F_{i},
\end{split}
\end{equation}
where $f_{i}^{eq}$ are the equilibrium distribution functions and $\tau$ is the relaxation time. 
The force term $F_{i}$ encodes the information of all the internal and external forces on the mixture.

The distribution functions are related to the total density $n$ and to the density momentum $n\textbf{u}$ through 
\begin{align}
\label{vel}
n&=\sum _{i}f_{i}, &  n\mathbf{u}&=\sum_{i}f_{i}\mathbf{e}_{i}+\frac{1}{2}\mathbf{F}\Delta t_{LB}
\end{align}
where $\textbf{F}$ is the force density to be properly determined. The expressions of the equilibrium distribution functions 
$f_{i}^{eq}$ must be chosen in order to locally conserve mass and momentum in each collision step. 
A convenient choice for the $D_{2}Q_{9}$ model is given by a second order expansion in the fluid velocity 
$\textbf{u}$ of the Maxwell-Boltzmann distribution
\begin{equation}
f_{i}^{eq}=\omega_{i}n\left [1+ \frac{\mathbf{e}_{i}\cdot\mathbf{u}}{c_{s}^{2}} + 
\frac{\mathbf{u}\mathbf{u}:(\mathbf{e}_{i}\mathbf{e}_{i}-c_{s}^{2}\mathbf{I})}{c_{s}^{4}}\right ],
\end{equation}
where $c_s=e/\sqrt{3}$ is the sound speed, $\textbf{I}$ is the unitary matrix and $\omega_{i}$ 
are coefficients with values $\omega_{0}=4/9$, $\omega_{i}=1/9$ for H and V directions, and $\omega_{i}=1/36$ for D directions.
The previous expansion is valid in the limit of small velocity $u << c_s$.
The forcing term $F_i$ can be written as a power series at the second order in the lattice velocities~\cite{47}
\begin{equation}
F_{i}=\left ( 1-\frac{\Delta t_{LB}}{2\tau} \right )\omega_{i}\left [ \frac{\mathbf{e}_{i}-\mathbf{u}}{c_{s}^{2}} 
+ \frac{\mathbf{e}_{i}\cdot \mathbf{u}}{c_{s}^{4}}\mathbf{e}_{i}\right ]\cdot \mathbf{F},
\end{equation}
where the force density is
\begin{eqnarray}
F_{\alpha}=-\partial_{\beta} P_{\alpha \beta}+\partial_{\alpha}(n c_s^2)&=&
\partial_{\alpha} (nc_{s}^{2}-p^{i})-\phi \partial_{\alpha} \mu_{\phi}-c \partial_{\alpha} \mu_c\nonumber\\
&=&-\phi \partial_{\alpha} \mu_{\phi}-c \partial_{\alpha} \mu_c.
\end{eqnarray}
This expression allows to recover the continuity and the Navier-Stokes equation in the continuum limit~\cite{46}.
Finally the bulk viscosity is $\zeta=\eta$, and the shear viscosity is
\begin{equation}
\eta=nc_{s}^{2}\Delta t_{LB}\left ( \frac{\tau}{\Delta t_{LB}} -\frac{1}{2}\right ).
\end{equation}
The implementation of bounding walls follows that described in Ref.~\cite{48}.

\subsection{Numerical solution of the convection-diffusion equations}

The convection-diffusion equations are solved by using a finite-difference scheme~\cite{46}.
Time is discretized in time steps $\Delta t_{FD}$ with time values given by $t^{n}=n\Delta t_{FD}$ ($n=1,2,3,...$). 
Throughout our simulations we have $\Delta t_{FD} =\Delta t_{LB}$. Any discretized function at time $t^{n}$ at the node $(x_{i},y_{i})$ is indicated 
with the symbol $g_{ij}^{n}$. To enhance the numerical stability, the finite-difference scheme is implemented by using two steps.
In the first step, the concentration field $\phi^{n}$ and the surfactant $c^n$ are updated in an intermediate time using an Euler algorithm
\begin{equation}
\phi^{n+1/2}=\phi^{n}-\Delta t_{FD}(\phi^{n}\partial_{\alpha}u_{\alpha}^{n}+u_{\alpha}^{n}\partial_{\alpha}\phi^{n})
\end{equation}
\begin{equation}
c^{n+1/2}=c^{n}-\Delta t_{FD}(c^{n}\partial_{\alpha}u_{\alpha}^{n}+u_{\alpha}^{n}\partial_{\alpha}c^{n}),
\end{equation}
where space derivatives of velocity are computed by using a centered scheme
while those of the concentration are calculated with an upwind scheme~\cite{46}.
The diffusive parts are implemented in the second step, as 
\begin{eqnarray}
&&\phi^{n+1}=\phi^{n+1/2}+\Delta t_{FD} \nabla^2 [a \phi^{n+1/2}+b f^{n}-\nonumber\\
&&(B_{0}+B_{1}c^{n+1/2})\nabla^{2}\phi^{n+1/2}-B_{1}\nabla c^{n+1/2}\cdot \nabla\phi^{n+1/2}],\nonumber\\
\end{eqnarray}
and
\begin{equation}
\begin{split}
c^{n+1}=c^{n+1/2}+\Delta t_{FD}[ \nabla \cdot ( Dc^{n+1/2} \nabla\mu_{c}^{n+1/2})\\
-\gamma(c^{n+1/2}-c_{0})+A\Theta(R-|\mathbf{r}-\mathbf{r}_{P}|) ],
\end{split}	
\end{equation}
where $f^{n}=(\phi^{n})^{3}$. Here spatial derivatives are discretized by using a centered scheme.

At walls neutral wetting boundary conditions are chosen by setting
$\mathbf{n} \cdot \nabla \mu_{\phi}=\mathbf{n} \cdot \nabla \mu_c=0$,
where $\mathbf{n}$ is a unit vector normal to the walls and pointing inwards.

\section{Equilibrium properties}

In order to validate our model, we first study the relaxation towards equilibrium of a planar sharp interface separating the phases of a binary fluid mixture. The study is performed on
a lattice of size $L_{x}=L_{y}=128\Delta x$ where periodic boundary conditions are set. Thermodynamic parameters are $a=-b=-0.001$, $D=0.1$, $\Delta x=\Delta t_{LB}=1$, $\tau=\Delta t_{LB}$.
We have also set $\gamma=A=0$ to neglect production and consumption contributions of the surfactant. A mapping with real units is reported in Appendix.
In Fig.~\ref{fig1} we show the profile of $\phi$ and $c$ at several simulation times, 
when $B_{0}=0.006$ and $B_{1}=-0.05$, in order to favour the migration of the surfactant (initially set to $c_i=0.1$ everywhere) to the interface from the bulk~\cite{49}.
\begin{figure*}
\begin{center}
\subfigure[]{
\centering
\includegraphics[width=0.5\textwidth]{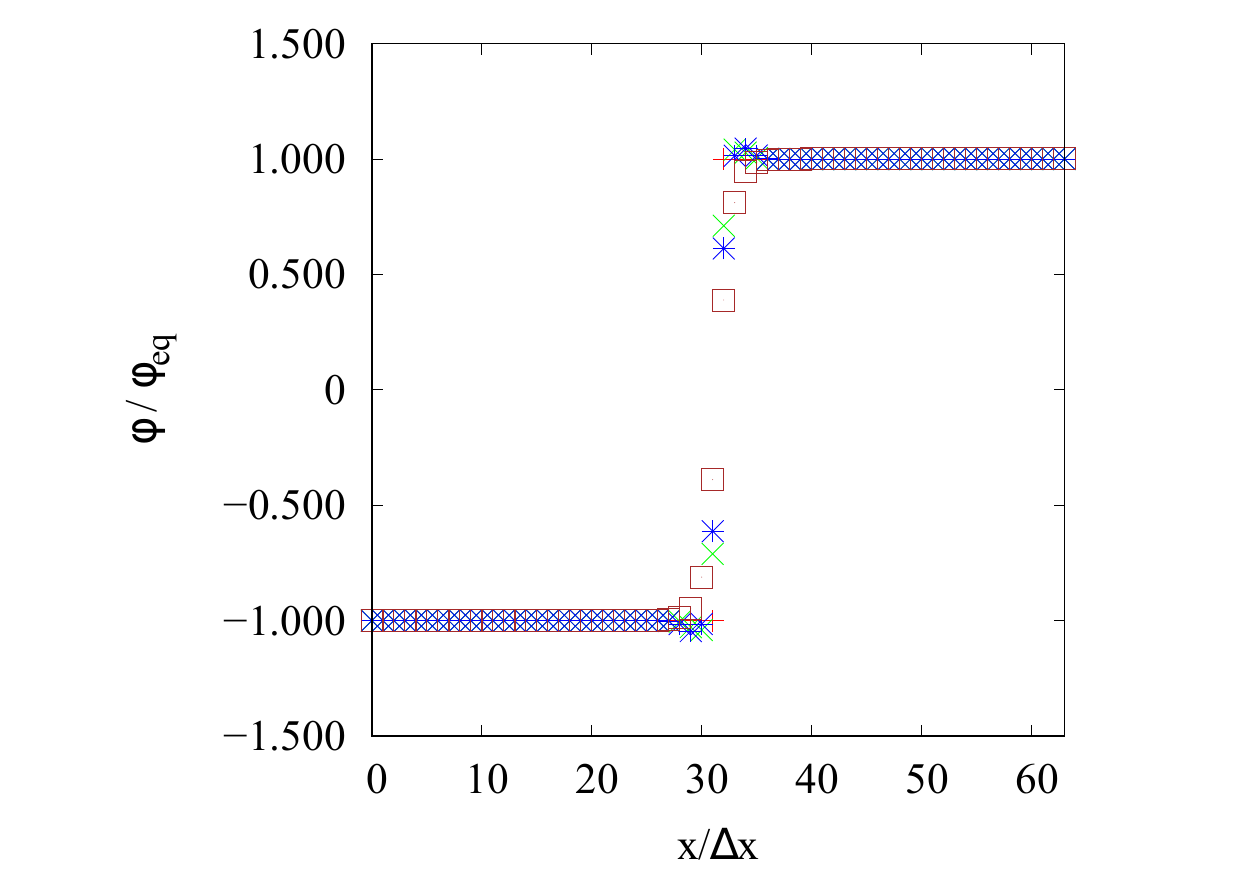}
}\\ 
\subfigure[]{
\centering
\includegraphics[width=0.5\textwidth]{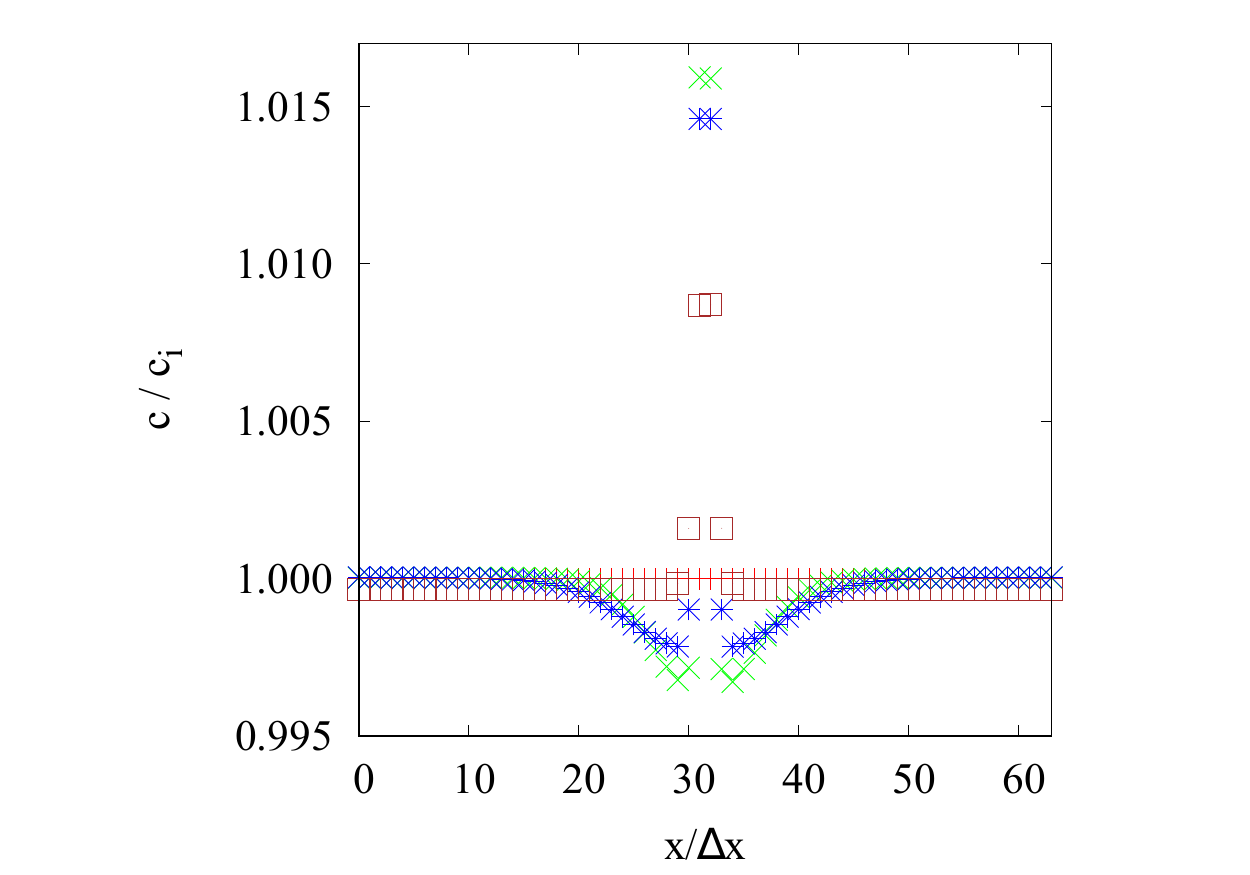}
}
\end{center}
\caption{Concentration profiles of $\phi$ (a) and $c$ (b) across the system for $B_{0}=0.006$ and $B_{1}=-0.05$ at times $t/\Delta t_{LB}=0$ ($+$, red), 
$10^{2}$ ($\times$, green), $2\times10^{2}$ ($\ast$, blue), $2\times10^{5}$ ($\Box$, brown).}\label{fig1}
\end{figure*}

The concentration $\phi$ correctly relaxes to its expected equilibrium profile, given by~\cite{41}
\begin{equation}
\phi(x)=\phi^{eq} \tanh \left(\frac{2x}{\xi}\right)
\end{equation}
where $\xi$ is the interface width, whereas the surfactant gradually moves towards the interface where a peak in the concentration is found. 
This is created at early times and lasts until full equilibration (namely when $\nabla\mu_{\phi}=\nabla\mu_{c}=0$ and $\partial_{\alpha}P_{\alpha\beta}=0$) is achieved. 
In line with these results, a good agreement is also found between the theoretical and the numerical values of the surface tension obtained for several values of $B_{0}$ and $B_{1}$ (see Table 1). 
For a planar interface along the $x$ direction, the former is calculated as~\cite{41}
\begin{equation}
	\label{Surf.1}
	\sigma_{th}=\int B(c)\left(\frac{\partial \phi}{\partial x}\right)dx,
\end{equation}
whereas the numerical value is computed from 
\begin{equation}
	\label{Surf.2}
	\sigma_{num}=\int (f(\phi,c)-f(\phi^{eq},c))dx,
\end{equation}
where $f(\phi,c)$ is the free-energy density of a system with planar interface and $f(\phi^{eq},c)$ is the free-energy density of 
a homogeneous system without interface.
\begin{table}
\caption{Comparison of theoretical and numerical values of the surface tension for a planar interface with different values of $B_0$ and $B_1$.}\label{tab:1}    \begin{tabular}{llll}
\hline\noalign{\smallskip}
{$B_{0}$} & {$B_{1}$} & {$\sigma_{th}$} & {$\sigma_{num}$} \\
\noalign{\smallskip}\hline\noalign{\smallskip}
0.003 & -0.01 & 0.00128 & 0.00129\\
0.003 & 0.01  & 0.00185 & 0.001856\\
0.006 & -0.01 & 0.00208 & 0.00208\\
0.006 & 0.01  & 0.00248 & 0.00249\\
0.006 & -0.03 & 0.00159 & 0.00160\\
0.006 & 0.03  & 0.00283 & 0.00284\\    
0.006 & -0.05 & 0.00086 & 0.00086\\
0.006 & 0.05  & 0.00314 & 0.00314\\
0.01 & -0.01  & 0.00283 & 0.00283\\
0.01 & -0.05  & 0.00208 & 0.00208\\
0.01 & -0.08  & 0.00127 & 0.00128\\
0.08 & -0.05  & 0.00915 & 0.00932\\ 
0.08 & -0.1   & 0.0089  & 0.0089\\
0.08 & -0.6   & 0.0043   & 0.0041\\		
\end{tabular}
\end{table}
By keeping $B_0$ fixed, the value of the surface tension is lower when $B_{1}<0$ due to the accumulation of surfactant at the interface.
As expected, by increasing $B_0$ the surface tension augments if $B_1$ is kept negative.

\begin{figure*}
\begin{center}
\includegraphics[width=8cm]{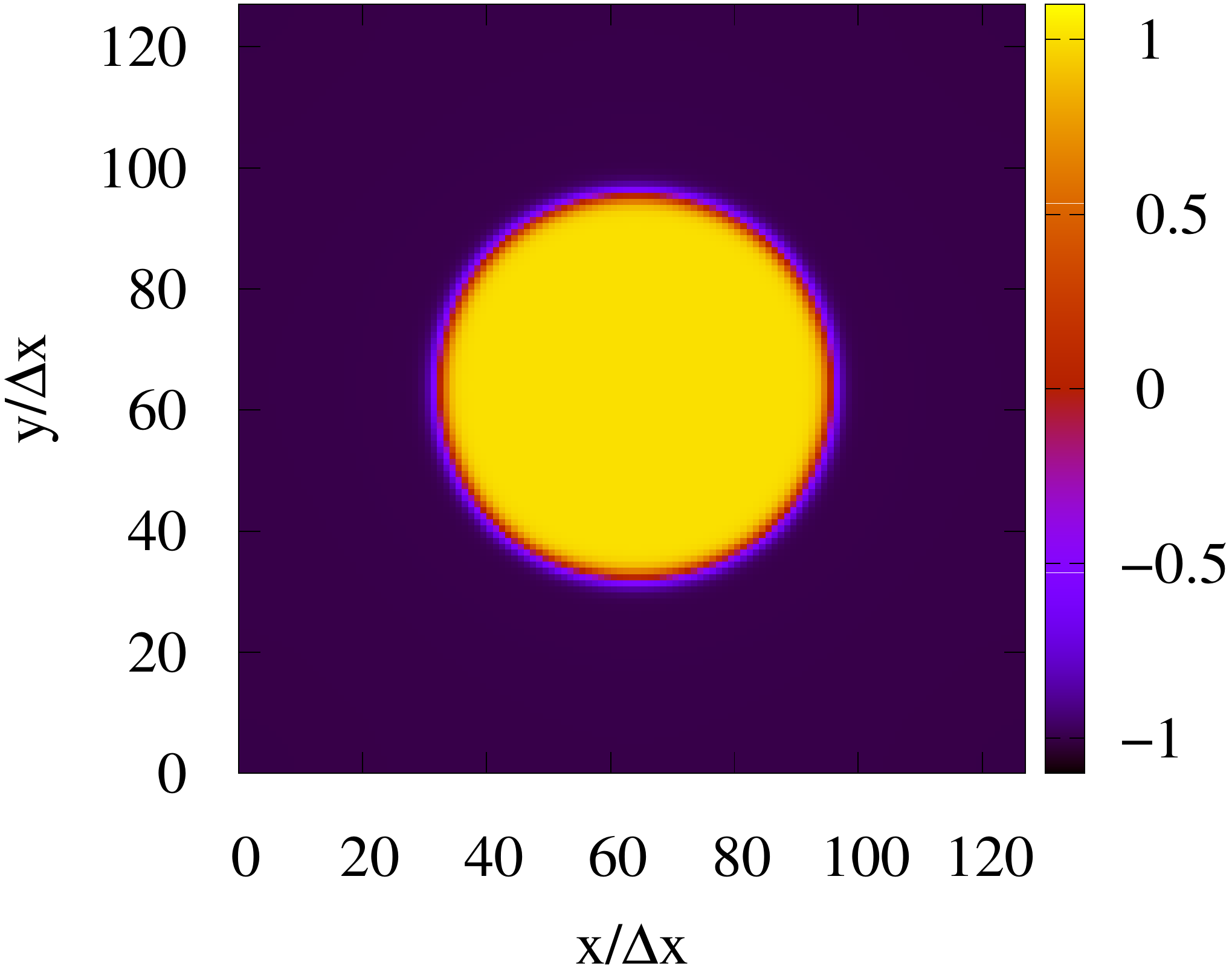}
\end{center}
\caption{Contour plot of $\phi$ at equilibrium for $B_{0}=0.006$ and $B_{1}=-0.05$. The red circle represents the area where the surfactant migrates. The colour scale for the contour plot also applies to Fig. 6, 7, 8 and 9, and refers to the values of $\phi/\phi_{eq}$.}\label{fig2}
\end{figure*}

We then studied the relaxation towards equilibrium of a circular droplet of radius $R=32\Delta x$ with the same parameter set, as a test case of a system with interface not aligned with lattice links.
Fig.~\ref{fig2} shows the contour plot of the concentrations $\phi$ of the system. No effects of the underlying lattice can be seen in the concentration configurations while the surfactant migrates to the interface. By using Eq.~(\ref{Surf.1}), we have also computed the surface tension of the droplet for some values of $B_0$ and $B_1$, considering a cross-section along the $x$-direction. 
Values reported in Table~\ref{tab:2} are in very good agreement with those obtained for a planar interface listed in Table~\ref{tab:1}. Same results have been obtained for a cross section taken along the 
$y$-direction. Finally, since the equilibrium relaxation time of a droplet depends upon its radius $R$, the viscosity of the fluid $\eta$ and the surface tension $\sigma$, a typical relaxation time
can be defined as $\tau_r=\eta R/\sigma$. Simulations of equilibrated droplets with different radii show that the numerical relaxation time scales linearly with $R$. In particular we have
$\tau_r/\Delta t_{LB}=3.11\times 10^4, 4.64\times 10^4, 6.22\times 10^4$ for $R/\Delta x=16,24,32$ respectively.

\begin{table}
\caption{Calculation of numerical values of the surface tension for a droplet, considering a cross-section along the $x$-direction.}\label{tab:2}
\begin{tabular}{lll}
\hline\noalign{\smallskip}
{$B_{0}$} & {$B_{1}$} & {$\sigma_{th}$} \\
\noalign{\smallskip}\hline\noalign{\smallskip}
0.006 & -0.03 & 0.00159\\
0.006 & 0.03  & 0.00281\\    
0.006 & -0.05 & 0.00086\\
0.006 & 0.05  & 0.00313\\
\end{tabular}
\end{table}

\section{Self-propelled droplet}
In this section we study the dynamics of an isolated self-propelled droplet surrounded by a passive Newtonian fluid.
We begin by considering a droplet of radius  $R=32\Delta x$ located at the centre of a relatively long rectangular box, with dimensions $L_{x}=512\Delta x$ and $L_{y}=256\Delta x$, and sandwitched between 
two plates at separation $L_y$.
The lattice length $L_x$ is chosen in order to minimize the artificial interaction between periodic images of the droplet. The interaction of the droplet with the walls can be considered negligible if 
the distance between the droplet interface and the wall itself is larger than $80\Delta x$, an empirical value estimated from a series of simulations performed at several vertical distances.
The concentration of the surfactant is initially constant all over the system and sets equal to $c_{i}=0.1$.
We have also set $B_{0}=0.006$ and $B_{1}=\pm0.05$, while the relaxation time is $\tau=0.6\Delta t_{LB}$.

\begin{figure*}
\begin{center}
\includegraphics[width=10.5cm]{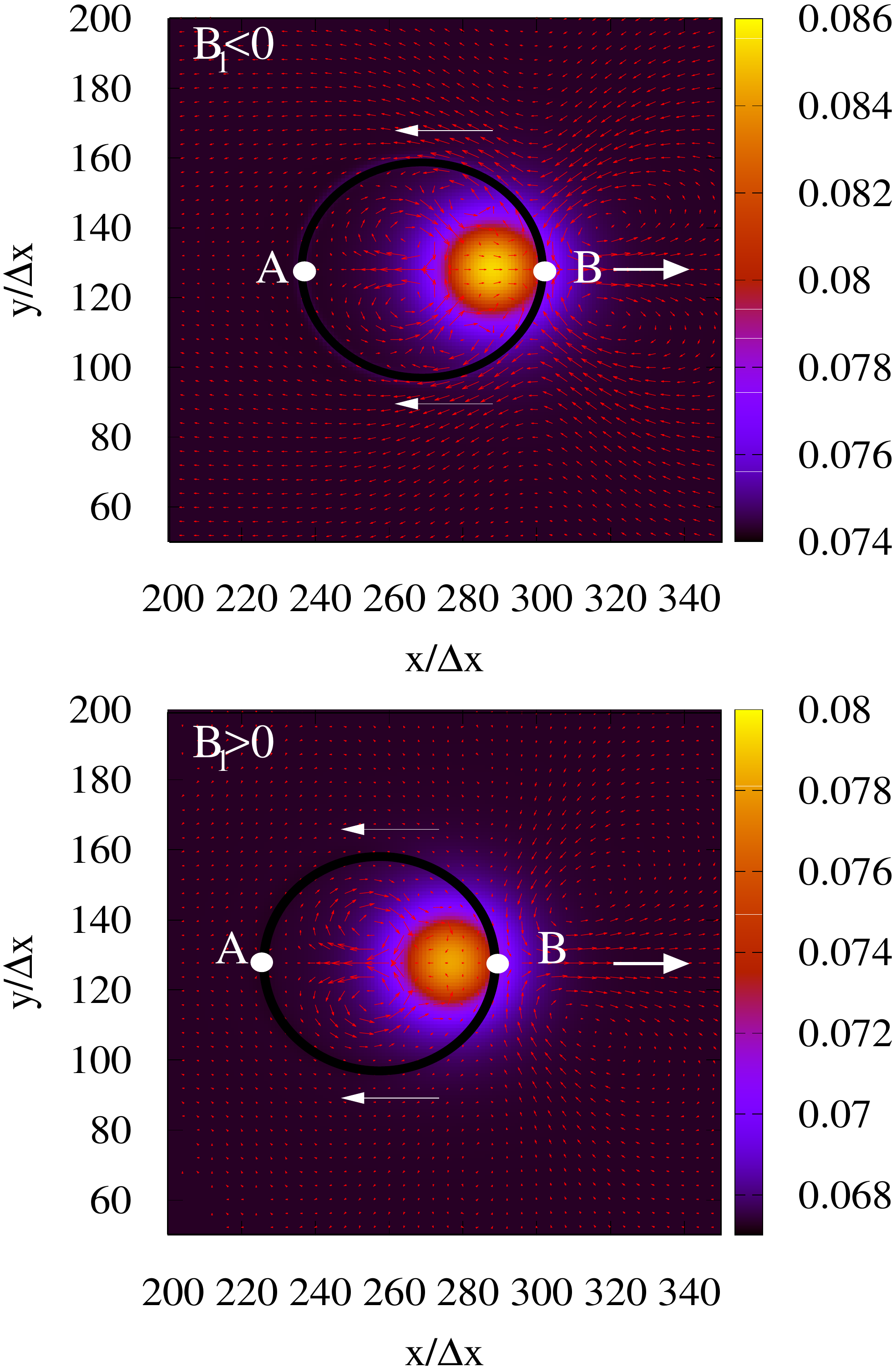}
\end{center}
\caption{Contour plot of the concentration field $c$ of the surfactant and of the velocity field (in the moving frame of the droplet) when $B_1<0$ (top panel) and $B_1>0$ (bottom panel)
at $t/\Delta t_{LB}=6\times10^{5}$ in both cases. The production and the consumption of surfactant starts at $t/\Delta t_{LB}=2\times10^{5}$. The black external circle indicates the interface of the droplet
at $\phi=0$ while the small circular yellow region is that in which the surfactant is produced. $A$ and $B$ indicate two extremal points of the droplet, with surface tension in $A$ greater than in $B$. 
The thin arrows on the top and on the bottom of the droplet indicate the direction of the Marangoni-like flow coincident with the gradient of surface tension, 
instead the thick arrow on the right of the droplet indicates the direction of motion. Only the central part of the rectangular simulation box is shown. The color scale refers to the values of $c$.}
\label{fig3}
\end{figure*}

We initially equilibrated the system until approximately $t/\Delta t_{LB}=2\times10^{5}$, when the total free energy is at its global minumum and the fluid mixture is fully equilibrated. Afterwards we switched on consumption and production terms (namely $A$ and $\gamma$) of surfactant in Eq.~(\ref{Equ.Surf.}). In our simplified model the concentration of surfactant is consumed at constant rate $\gamma=1.5 \times 10^{-3}$ in the whole system, while its production occurs in a circular region inside the droplet and tangential to it (see Fig.~\ref{fig3}). 
In particular the coefficient $A$ (that controls the strength of production) was chosen time dependent with $A(t)=A_{0}t$ 
for $t<t^*=2\times 10^6\Delta t_{LB}$, and was kept constant to the value $\simeq 10^{-4}\div 10^{-5}$ for $t\ge t^*$. 
The value of $\gamma$ and the functional form of $A$ guarantee a good numerical stability (as long as $\gamma/A_0\simeq 10^7$) and a considerable movement of the droplet. 
Indeed if the production of surfactant is too fast, the stability condition $B(c)>0$ is violated and the droplet breaks up.

\begin{figure*}
\begin{center}
\includegraphics[width=1\textwidth]{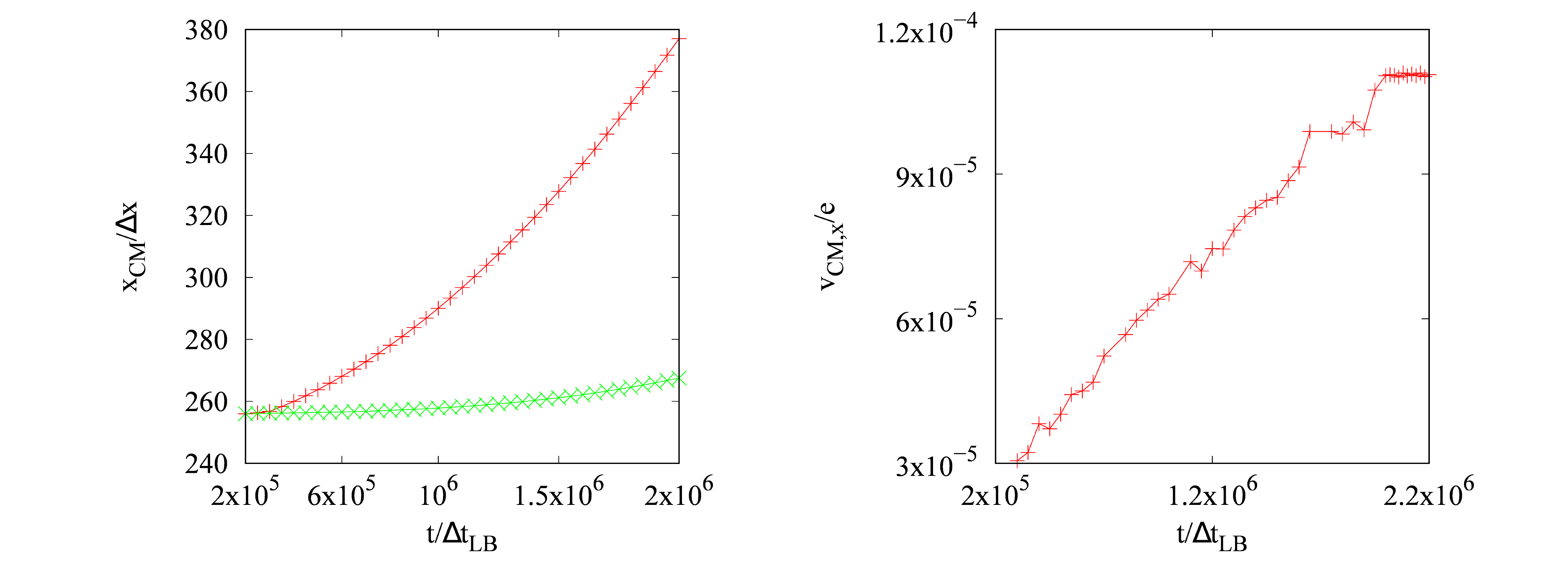}
\end{center}
\caption{(Left) $x$ component of the center of mass of the droplet as a function of time for $B_{1}=-0.05$ ($+$, red) and for $B_{1}=0.05$ ($\times$, green) in the time interval 
$[2 \times 10^{5}\Delta t_{LB}; 2 \times 10^{6}\Delta t_{LB}]$. The droplet starts from the middle of the lattice ($L_{x}/2=256\Delta x$, $L_{y}/2=128\Delta x$) and moves unidirectionally rightwards 
along the increasing abscissa. (Right) $x$ component of center of mass velocity  for $B_1=-0.05$ in the same time interval. Velocity increases almost linearly up to $t^*/\Delta t_{LB}=2\times 10^{6}$,
until the production term is on. For $t\ge t^*$ $A(t)$ is switched off and the velocity remains constant.}
\label{fig4}
\end{figure*}

After consumption and production terms are switched on, the droplet acquires motion mainly parallel to the $x$-axis along the increasing abscissa, regardless of the sign of the coefficient 
$B_1$ (see Fig.~\ref{fig3}). This occurs as a lower surface tension, created at the interface where the surfactant is produced, triggers the formation of a Marangoni-like
flow favouring the onset of the motion. However the velocity field structure is strongly affected by the sign of $B_1$. Indeed if $B_1<0$ (top panel, Fig.~\ref{fig3})
a quadrupolar-like fluid flow forms inside the droplet. Two vortices form mainly tangentially to the interface (the top one rotating counterclockwise while the bottom one rotating clockwise) 
and two further vortices are created in the bulk of the droplet (the top one rotating clockwise and the bottom one rotating counterclockwise). The local fluid is pushed backwards mainly along the interface
near the leading edge (point B in Fig.~\ref{fig3})
and from the centre of the droplet (in direction antiparallel to the $x$-axis) whereas it is pumped forwards diagonally in the region where the surfactant is produced. Outside the droplet instead, 
two large vortices form at its leading edge, favouring the motion along the increasing $x$-axis. If $B_1>0$ instead, there is less surfactant on the interface and more in the bulk of the droplet; besides
the two vortices formed on the right outside it, only two further large vortices (the upper one rotating clockwise and the lower one rotating counterclockwise) are created inside,
with a negligible fluid flow near the interface. We also note that when $B_1<0$ the magnitude of the velocity field is more intense than when $B_1>0$, hence the droplet
moves faster. This can be seen by looking at the position of the center of mass of both droplets, computed as $\mathbf{r}_{CM}(t)=\sum_{ij}\phi_{ij}\mathbf{r}_{ij}(t)/\sum_{ij}\phi_{ij}$
where the indexes $i$ and $j$ denote the lattice sites and the sum is calculated under the constraint $\phi_{ij}\geq 0$.
Fig.~\ref{fig4} shows the evolution of the $x$-component of the centre of mass in the time interval 
$[2 \times 10^{5}\Delta t_{LB} ; 2 \times 10^{6}\Delta t_{LB}]$ in which both droplets move unidirectionally but with a different speed, higher when $B_1<0$.
We finally mention that, when consumption and production are switched on, the surfactant concentration in the bulk reduces to $c\simeq 0.06$ if $B_1<0$
(its complete depletion is prevented by the existence of a minimum at a finite value of $c>0$ in the bulk free energy term $c\ln c$),
a value significantly lower than that at the interface.
The surfactant concentration is large only in a very narrow area in front of the droplet which extends axially over a distance up to $R/3$.

\begin{figure*}
\begin{center}
\subfigure[]{
\centering
\includegraphics[width=0.5\textwidth]{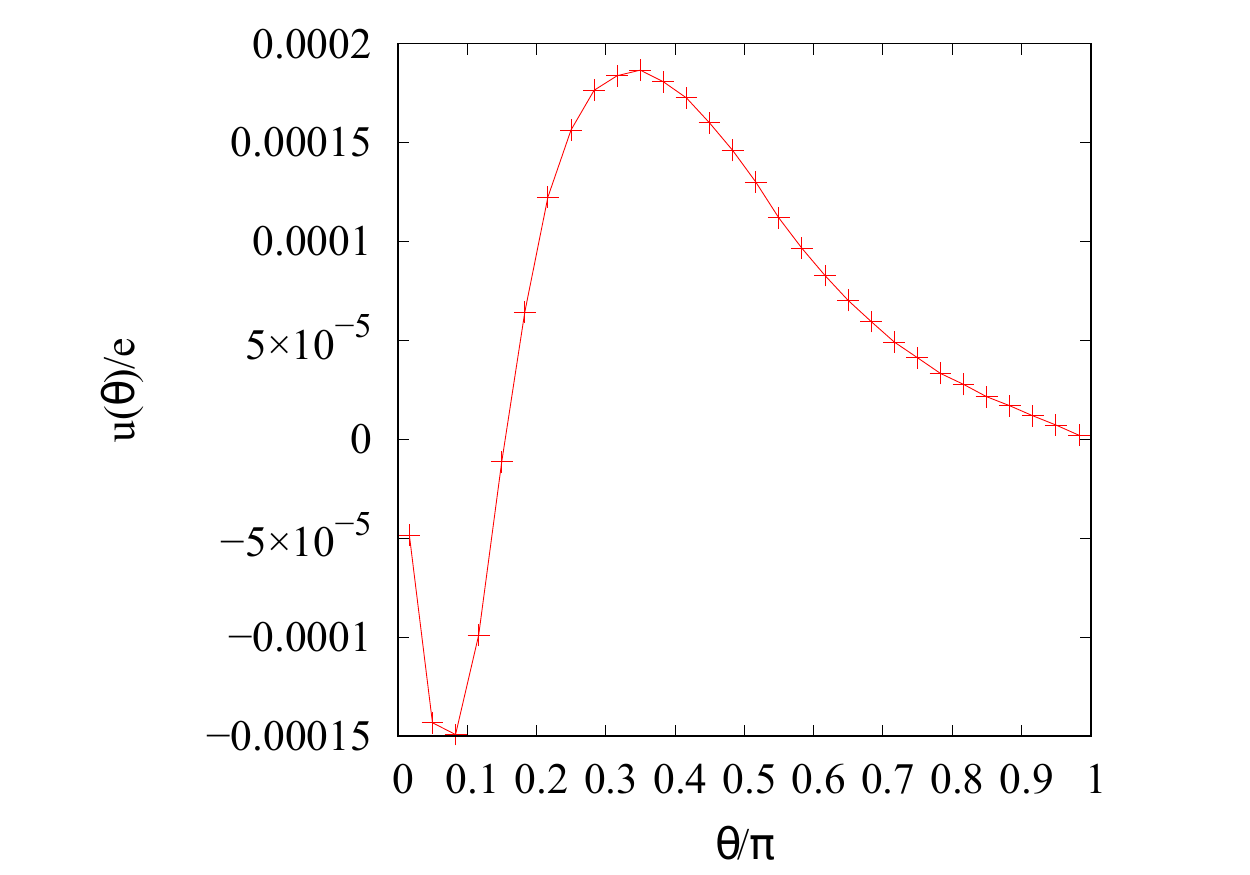}}
\subfigure[]{
\centering
\includegraphics[width=0.5\textwidth]{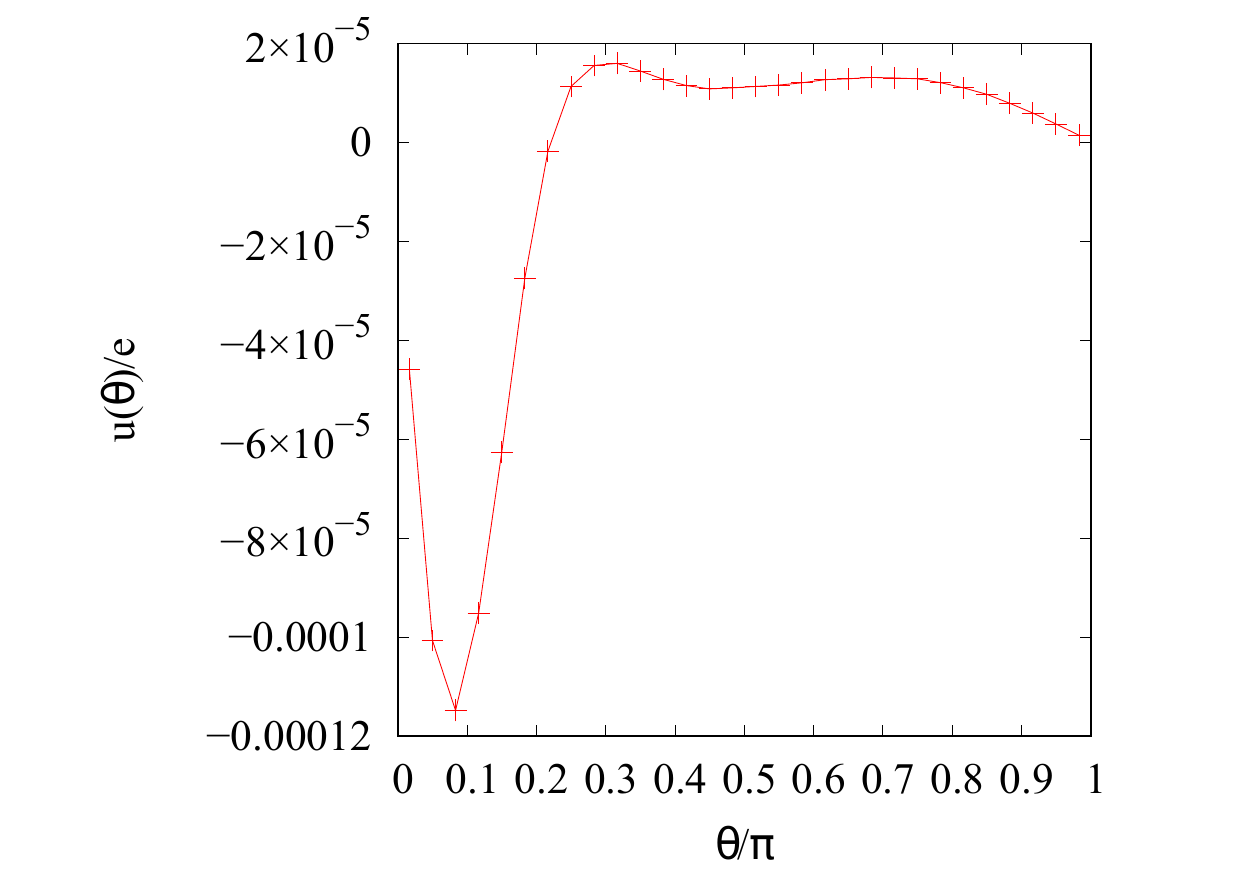}}
\end{center}
\caption{Polar velocity $u({\theta})$ as a function of the polar angle $\theta$ for $B_{1}<0$ (a) and $B_{1}>0$ (b). Data are taken at time $t/\Delta t_{LB}=6 \times 10^{5}$.}\label{fig5}
\end{figure*}

The droplet described so far is an example of an active system as it is capable to perform autonomous motion by converting energy of a chemical source, such as that stemming from
an inhomogeneous production of surfactant. The dynamics of such active swimmers can be theoretically modeled by the
``squirmers'', spherical particles with a predefined axisymmetric tangential velocity
distribution on its surface whose deformation can trigger motion~\cite{10}. 
Depending on the structure of the velocity field a single squirmer creates in the surrounding fluid, three classes can be distinguished, namely pushers (or extensile swimmers), pullers (or contractile
swimmers) and neutral swimmers. In the particle frame, the fluid will be pulled inwards equatorially and emitted axially for pushers whereas the opposite will occur for pullers. 
While in both cases the velocity field is dipolar, that associated to a neutral swimmer is quadrupolar~\cite{51}.
In the lab frame instead, the propulsion will act from the rear for pushers and from the front for pullers with a characteristic double-vortex structure of the velocity field.
The typical velocity profile on the surface of a squirmer can be written as~\cite{51,53}
\begin{equation}\label{eq_v}
u(\theta)=b_{1}sin(\theta)+\frac{1}{2}b_{2}sin(2\theta)
\end{equation}
where the two terms on the right hand side stem from an expansion with Legendre polynomials truncated to the second mode. 
Eq.~(\ref{eq_v}) represents the polar component of the velocity of the fluid and $\theta$ is the polar angle. $\theta=0$ defines the direction in which the squirmer swims, 
namely that indicated by the thick arrow in Fig.~\ref{fig3}. The dimensionless parameter $\beta=b_{2}/b_{1}$ defines the type of squirmer: if $\beta<0$ it is a pusher, if $\beta>0$ it is a puller and
if $\beta=0$ it is a neutral swimmer.
In order to determine the dynamical behaviour of our active droplet we calculated $u(\theta)$ from our numerical data for both $B_1<0$ (Fig.~\ref{fig5}a) and $B_1>0$ (Fig.~\ref{fig5}b) at
$t/\Delta t_{LB}=6 \times 10^{5}$ and fitted it by using 
Eq.~(\ref{eq_v}). Although only on a qualitative level, our results support the view that the droplet behaves as pusher-like particle (regardless of the sign of $B_1$),
as the propulsion acts mainly from the rear of the droplet~\cite{51,53,54}. 
A better agreement between the analytic velocity profile and the simulated one could require more complex theoretical analysis, such as a fit in which higher order terms in the Legendre polynomials are considered. We note however that the flow field profile in the front of the droplet is difficult to capture quantitatively due to its complicated coupling with the surfactant gradient (see Eq.~\ref{Equ.Surf.}).

\section{Scattering of two self-propelled droplets}

In this section we investigate the interaction between two droplets.
We consider different system realizations in which both active and passive droplets are present. The latter is modeled as a surfactant-free system in which an autonomous motion is inhibited.
In all the cases studied their interaction is mainly mediated by the velocity field, since they remain at a relative distance where interaction between surfactant of different droplets 
can be neglected. Simulations are shown when $B_1<0$ as, for the set of parameters considered, a more persistent motion is found.

\subsection{Colliding collinear droplets}

We first consider the case of two collinear droplets of radius $R_0 = 32 \Delta x$ located on the same horizontal line in a lattice of size $L_{x}=512\Delta x$, $L_{y}=256\Delta x$.
They are placed at distance 
$|x_{cm,1}(t_i)-x_{cm,2}(t_i)| = d_x(t_i) = 6R_{0}$, where $(x_{cm,1},y_{cm,1})$ and $(x_{cm,2},y_{cm,2})$ are the coordinates of the centers of mass
of the droplets. The same set of parameters used in the case of the isolated active droplet is adopted here as well.
The system is initially relaxed up to $t_i \simeq 2\times10^{5}\Delta t_{LB}$; afterwards production and consumption terms of the surfactant are switched on.
\begin{figure*}
\includegraphics[width=1.0\textwidth]{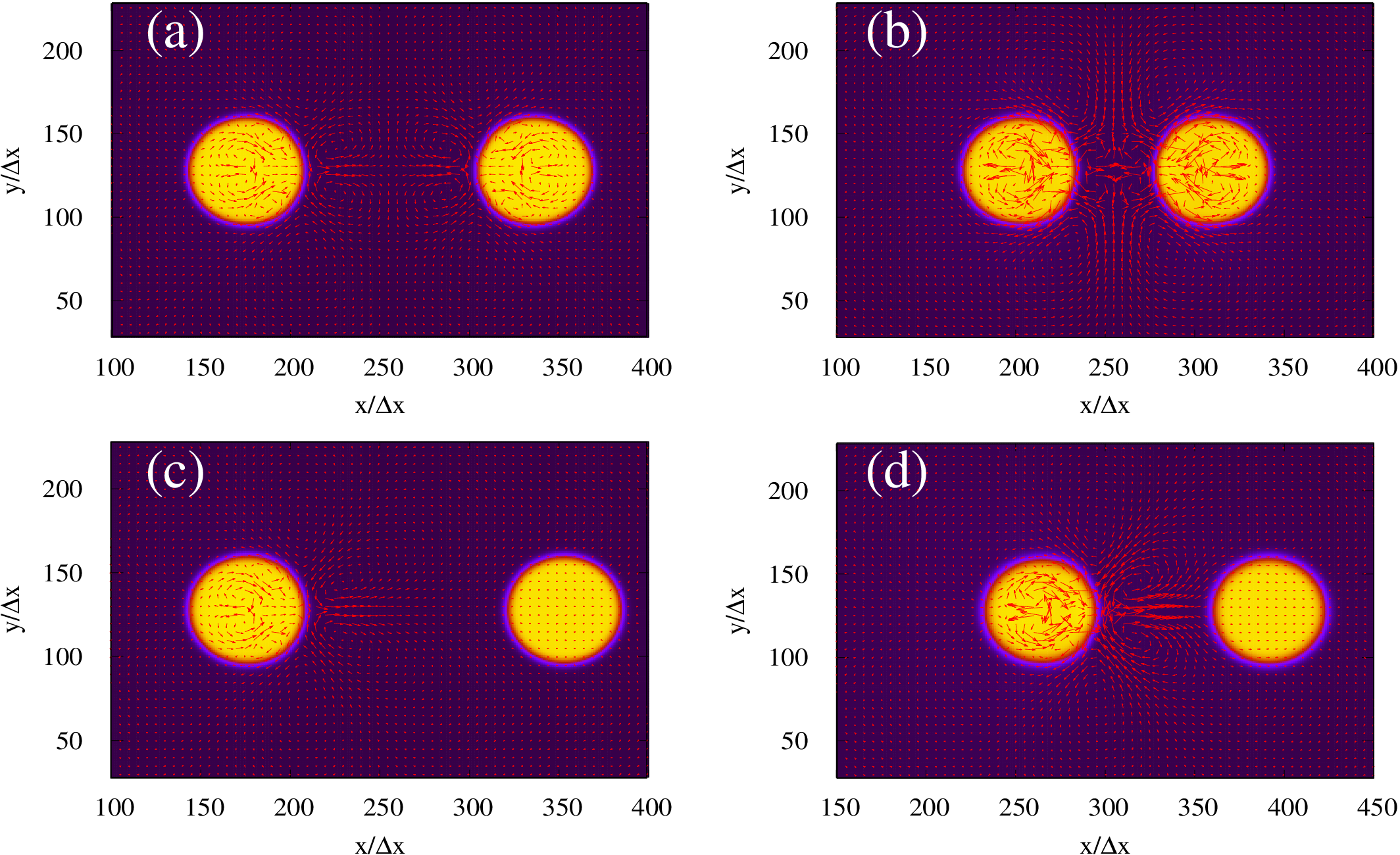}
\caption{Contour plots of the concentration field $\phi$ for two active colliding collinear droplets ((a)-(b)), and for an active left droplet moving towards a passive right droplet ((c)-(d)).
The velocity field is also shown. The whole lattice is $[512\Delta x \times 256\Delta x]$. In the first case the two active droplets initially approach each other ((a) $t/\Delta t_{LB}=7\times10^{5}$), 
but, when a balance between the repulsion generated by intermediate hydrodynamic flow and the attraction due to a Marangoni-like effect is achieved, the reciprocal motion ceases ((b), $t/\Delta t_{LB}=1.585\times10^{6}$). In the second case the active droplet moves towards the passive immobile one ((c), $t/\Delta t_{LB}=7\times10^{5}$) and, when they come sufficiently close, the former pushes the latter to the right ((d), $t/\Delta t_{LB}=2.2\times10^{6}$).}\label{fig6}     
\end{figure*}
At early times, the velocity field generated by each droplet resembles that seen in Fig.~\ref{fig3}, with four vortices inside the droplets and other four ones
located in their front, arranged into a fourfold symmetry. Due to the initial relatively large distance between them, the hydrodynamic interaction mediated by the external passive fluid
is negligible (Fig.~\ref{fig6}a). However, as the droplets move against each other driven by the inhomogeneity of the 
surfactant distributions, a compression of the four vortices formed in the fluid in between is observed. In addition a further quadrupolar field is clearly visible in the system. Here two vortices are located on top of each droplet outside them and the other two are placed at their bottom (Fig.~\ref{fig6}b). 
The structure of the velocity field inside each droplet instead displays the typical four vortices, again triggered by a Marangoni-like effect, although now more intense than that seen at early times. 
The relative motion persists up to a time $t_{f}$ when a balance between the repulsion induced by the flow between the droplets and 
an attraction due to Marangoni effect is achieved. This occurs, for the specific simulation considered, at distance $d_x(t_f) \simeq 3.3R_{0}$
with $t_f/ \Delta t_{LB} \simeq 1.585\times 10^{6}$.

The dynamics is overall simpler if the active droplet on the right is replaced with a passive one (Fig.~\ref{fig6}c-d).
In this case only the active droplet on the left is self-propelled along the increasing $x$-axis. While initially the interaction between the two
droplets is negligible (Fig.~\ref{fig6}c), when the active one approaches the passive one the fluid flow generated by the former pushes the latter rightwards,
which is hence simply advected by the fluid (Fig.~\ref{fig6}d).
Finally in the absence of surfactant both droplets would be passive and they would either remain firm in equilibrium or coalesce if close enough.

Hence even in a relatively simple realization of a system with two droplets, 
the dynamics can be dramatically different 
and ultimately be affected by the nature of the droplets considered. In particular an inhomogeneous production of surfactant migrating
towards the interface of the droplet may either drive the system to a non-motile steady state (if both droplets are active) or render
motile a passive droplet if this is close enough with an active one. 
 
\subsection{Colliding non collinear droplets}

\begin{figure*}
\includegraphics[width=1.0\textwidth]{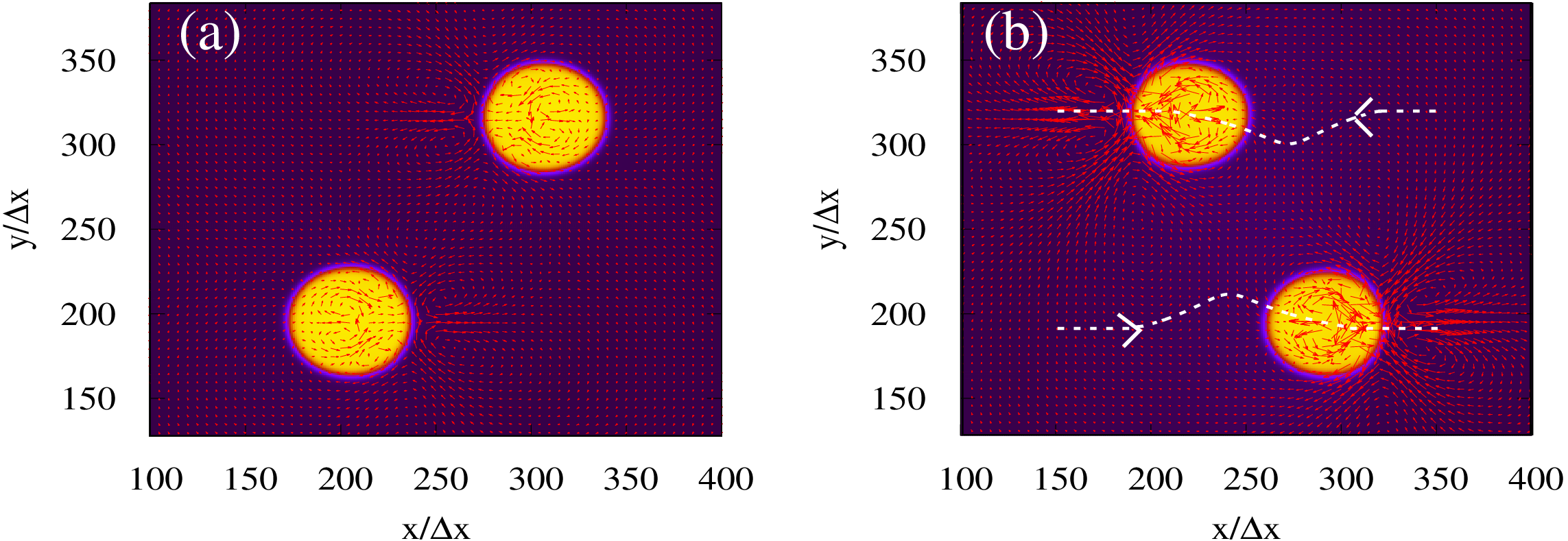}
\caption{Contour plots of the concentration field $\phi$ for two active vertically shifted colliding droplets. The velocity field is also shown. The configurations are at times $t/\Delta t_{LB}=7\times10^{5}$ (a) and $t/\Delta t_{LB}=2\times10^{6}$ (b) and the trajectory is also shown. The whole lattice is $L_{x}=512\Delta x \times L_{y}=512\Delta x$. The two active droplets are initially attracted by each other ((a), $t/\Delta t_{LB}=7\times10^{5}$) and, when close enough, they are scattered away along their own direction. The color scale is the same as in 
Fig.~\ref{fig2}.}
\label{fig7}
\end{figure*}

The dynamics is even more intriguing in a modified setup in which the droplets are vertically shifted. 
More specifically, two droplets of radius $R_0=32\Delta x$ are placed at $d_x(t_i) = 6.25R_{0}$ and at 
$d_y(t_i) = 4R_{0}$, in a lattice of size $L_{x}=512\Delta x$, $L_{y}=512\Delta x$ (Fig.~\ref{fig7}a).

We initially consider a realization in which both droplets are active, the upper one moving leftwards and the lower one moving 
rightwards (Fig.~\ref{fig7}). While at an early stage the velocity field formed in the fluid between the droplets (along a direction 
joining their centers of mass) favours their mutual attraction (Fig.~\ref{fig7}a), later on (see the trajectories in Fig.~\ref{fig7}b), 
after the droplets achieve a minimal vertical 
distance (estimated roughly $d_{y,min} \simeq 3R_{0}$), it induces a reciprocal repulsion, as clearly visible from the 
direction of the velocity field formed between the droplets when they pass each other (Fig.~\ref{fig7}b). 
After the scatter occurs, the droplets move along their own direction at an almost constant speed.
We remind that no rotation of droplets occurs since
the region of surfactant production is fixed inside the droplet and
does not follow its direction of motion.
This behavior has been observed for values of vertical separation $3.5R_{0}\leq d_y(t_i) \leq 4R_{0}$, whereas
for lower values the attraction is strong enough to overcome
the resistance to coalescence mediated by the surfactant.

\begin{figure*}
\includegraphics[width=1.0\textwidth]{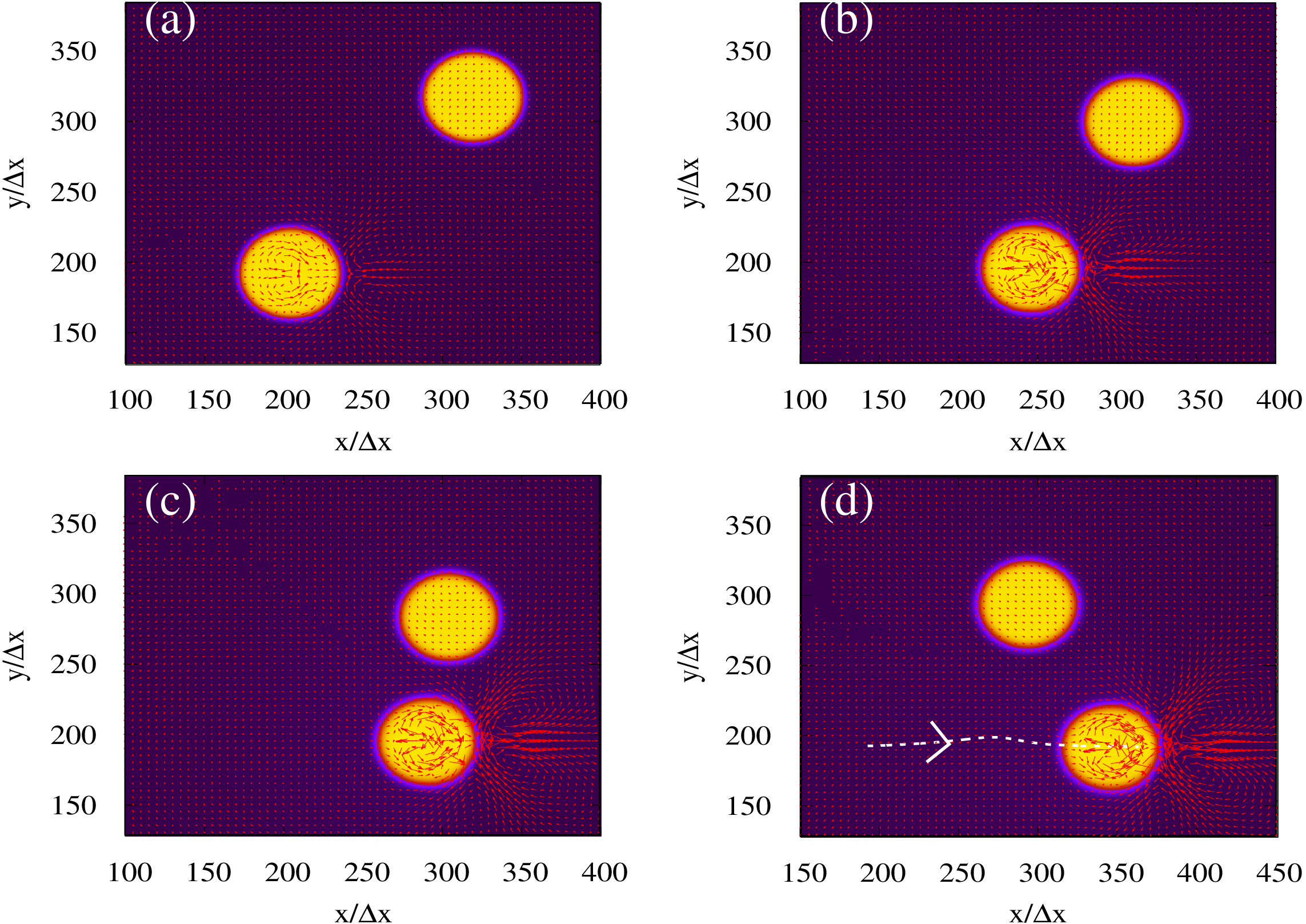}
\caption{Contour plots of the concentration field $\phi$ for an active and a passive vertically shifted colliding droplets. The velocity field is also shown. The configurations are taken at times $t/\Delta t_{LB}=7\times10^{5}$ (a), $t/\Delta t_{LB}=1.6\times10^{6}$ (b), $t/\Delta t_{LB}=2.3\times10^{6}$ (c) and $t/\Delta t_{LB}=3\times10^{6}$ (d).  The lattice size is $L_{x}=512\Delta x$, $L_{y}=512\Delta x$. The left active droplet moves towards the passive one ((a)-(b)) and when close enough, it initially attracts and then repels the passive one ((c)-(d)). Later on, the active droplet continues its  motion to the right 
along the dashed trajectory, leaving the passive droplet on its left (d). The color scale is the same as in Fig.~\ref{fig2}.}
\label{fig8}
\end{figure*}

If the upper active droplet is replaced with a passive one, the dynamics is not too dissimilar from the previous case.
Here the active droplet on the left acquires motion rightwards and, when close enough with the passive one, attracts it 
downwards (Fig. ~\ref{fig8} (a)-(c)), until a minimal vertical separation of $d_{y,min} \simeq 2.8R_{0}$ is achieved. 
Later on, when the two droplets are almost vertically aligned, the velocity field created by the active droplet
repels the passive one, which slightly shifts leftwards, whereas the active one continues its motion rightwards, 
along the dashed trajectory indicated in Fig.~\ref{fig8}d. The main difference with the case discussed in Fig.~\ref{fig7} is that
the droplets get even closer although at much longer times, as a weaker mutual interaction is now due to only one source.
Also in this case this behavior holds as long as $3.5R_{0} \leq d_y(t_i) \leq 4R_{0}$, whereas for lower values droplets coalesce.

\subsection{Vertically aligned droplets}

\begin{figure*}
\includegraphics[width=1.0\textwidth]{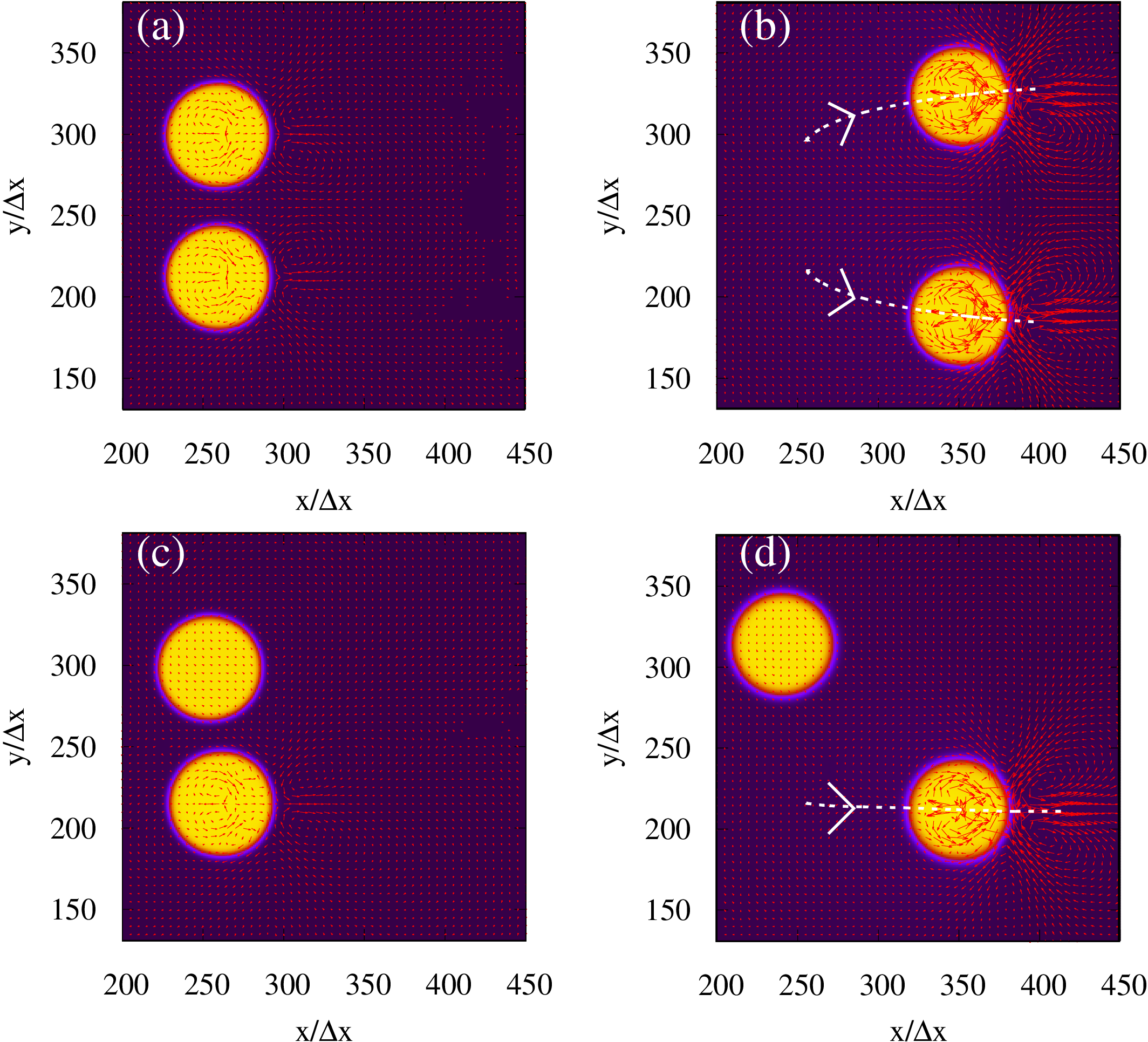}
\caption{Contour plots of the concentration field $\phi$ for two active equiverse droplets ((a)-(b) and active lower droplet and passive upper droplet ((c)-(d)). The velocity field is also shown. The lattice size is $L_{x}=512\Delta x \times L_{y}=512\Delta x$. In the first case the two active droplets initially repel each other and move rightwards ((a), $t/\Delta t_{LB}=5\times10^{5}$) along the direction indicated by the dashed lines ((b), $t/\Delta t_{LB}=2\times10^{6}$). In the second case the active lower droplet repels the upper passive one, which now drifts backwards ((c), $t/\Delta t_{LB}=5\times10^{5}$), 
and moves rightwards as in the previous case ((d), $t/\Delta t_{LB}=2.35\times10^{6}$). The color scale is the same as in Fig.~\ref{fig2}.}
\label{fig9}
\end{figure*}

We finally consider the case of two droplets vertically aligned, with $d_y(t_i) = 2.5R_{0}$ and  $d_x(t_i) = 0$ (Fig.~\ref{fig9}).
If both droplets are active (Fig.~\ref{fig9}a-b), the velocity field formed in between favours their reciprocal repulsion and a drift along the vertical direction, 
whereas that generated inside each droplet and in their front drags them along the positive $x$-axis.  Droplets move initially slightly faster along the $x$-direction rather than along the $y$-one,
but when they get at a vertical distance of $d_y \simeq 4.2R_{0}$, the speed along the $y$-axis progressively diminishes as the repulsion becomes negligible.
Our geometry slightly resembles that studied by Gompper et. al.~\cite{52} in which the interaction between two pullers is investigated.

If the upper droplet is passive (Fig.~\ref{fig9}c-d), the flow field created by the active one first pushes it upwards and afterwards drags it slightly backwards. 
The motion of the active droplet proceeds almost unaltered rightwards along the trajectory indicated in Fig.~\ref{fig9}d, and the repulsion with the passive one becomes
negligible at $t \simeq 2.35\times 10^{6}\Delta t_{LB}$, when the vertical separation is roughly $d_y \simeq 3.2R_{0}$.

\section{Conclusions}

In this work we have studied the chemically-driven self-propelled motion of an isolated droplet with lattice Boltzmann simulations, and we have also investigated the interaction between self-propelled 
droplets considering several system realizations, such as those in which they are either both active or one of them is passive. 

An autonomous motion can be achieved when an inhomogeneous concentration of surfactant, produced in a region inside the droplet and consumed outside,  is created in the system.
This determines a gradient in the surface tension on the interface of the droplet which triggers the formation of a Marangoni-like flow favouring the motion. The advantage of our minimal description
relies on the fact that it can provide a full resolution of the hydrodynamic field in the system, in particular the intricate structure of the vortices inside the droplet and in the surrounding fluid,
by controlling a small number of free parameters, such as $A$ and $\gamma$ in Eq.~(\ref{Equ.Surf.}) and $B(c)$.
Although at an initial stage, our preliminary study on the interaction of two active droplets shows that the dynamics strongly depends both on the geometry considered and on the nature of the droplets. 
If both active, they initially attract and then repel each other, depending on the structure of the velocity field in the surrounding passive fluid which mediates the interaction. If, instead, one of
the droplet is passive, the latter is simply advected by fluid flow generated by the active one. 

Our results are a first step towards a deeper understanding of the physics of more complex systems, such as those in which the surfactant is allowed to rotate in order to drive the droplet motion along
a pre-established direction. This is going to be of paramount importance when a large number of droplets is considered. In this case the mutual interaction
is generally non-trivial and could drive the system towards novel emergent behaviours in which the collective motion of droplets plays a crucial role. A careful control on the dynamics of the droplets
could be fundamental in order to create targeted structures starting from a primordial soup of disordered droplets. Finally, it would be desirable to extend our study to a fully 3D system which,
despite being very demanding in terms of computational resources, is certainly of great interest for pratical purposes.

\section*{Appendix}

Here we provide an approximate correspondence between our simulation parameters and typical realistic numbers of an active droplet.
A mapping of the radius $R$ and of the velocity $v$ of the simulated droplets to the corresponding typical values of experiments ($R= 40\mu$m and $v=15\mu$$m/s$)~\cite{13},
leads to a space and time step equal to $\Delta x\simeq 1.25\times 10^{-6}$m and $\Delta t_{LB}\simeq 7.5\times 10^{-6}$s, respectively. Hence the consumption rate is $\gamma=200$$s^{-1}$.
Furthermore, matching the initial surfactant concentration $c_i$ to the experimental value of $50$$mMl^{-1}$~\cite{13} 
provides an estimate of the production rate $A\simeq 6.6\times 10^2$$mMl^{-1}s^{-1}$.

\section*{Author contribution statement}

All authors contributed equally to the paper.\\


\begin{thebibliography}{}
\bibitem{1} 
R. Seemann, J.-B. Fleury, and C. C. Maass, Eur. Phys. J. Special Topics \textbf{225}, 2227 (2016); C. Maas, C. Kruger, S. Herminghaus and C. Bahr, Annu. Rev. Condens. Matter Phys. (2016).
\bibitem{2}
M.C. Marchetti, J.F. Joanny, S. Ramaswamy, T.B. Liverpool, J. Prost, M. Rao, R.A. Simha, Rev. Mod. Phys. \textbf{85}, (2013).
\bibitem{3} 
H. Tanaka, T. Araki, Phys. Rev. Lett. \textbf{81}, 389 (1998); P. Poesio, G. Beretta, T. Thorsen, Phys. Rev. Lett. \textbf{103}, 064501 (2009);
T. Ban, T. Yamada, A. Aoyama, Y. Takagi, Y. Okano, Soft Matter \textbf{8}, 3908 (2012). 
\bibitem{4}
J. Zhang, Y. Yao, L. Sheng, J. Liu, Advanced Materials \textbf{27}, 2648 (2015).
\bibitem{6}
J. D. Murray, \textit{Mathematical Biology} Springer-Verlag, New York, 1989.
\bibitem{7}
K. I. Agladze, V. I. Krinsky, and A. M. Pertsov, Nature London \textbf{308}, 834 (1984).
\bibitem{8}
R. Phillips, J. Kondev, J.A. Theriot, \textit{Physical biology of the cells} Garland Science, NY (2008).
\bibitem{9}
E. Tjhung, D. Marenduzzo and M. Cates, PNAS \textbf{109}, 1238 (2012); E. Tjhung, M. Cates and D. Marenduzzo, PNAS \textbf{114}, 4631 (2017).
\bibitem{10}
M.J. Lighthill, Comm. on Pure and Appl. Math. \textbf{109}, 118 (1952); M.J. Lighthill, Mathematical biofluiddynamics, Regional conference series in applied mathematics, (1975);
J.R. Blake, J. Fluid Mech. (1971), \textbf{46} pp.199-208.
\bibitem{ehlers}
K. M. Ehlers, D. Samuel, H. C. Berg, R. Montgomery, PNAS \textbf{93}, 8340 (1996).
\bibitem{drescher}
K. Drescher, K. C. Leptos, I. Tuval, T. Ishikawa, T. J. Pedley, R. E. Goldstein, Phys. Rev. Lett. \textbf{102}, 168101 (2009).
\bibitem{13}
S. Thutupalli, R. Seemann and S. Herminghaus, New J. Phys. \textbf{13}, 073021 (2011).
\bibitem{14}
R.F. Probstein, \textit{Physicochemical hydrodynamics: An Introduction}, John Wiley and Sons, 1994; A.A. Nepomnyashchy, M.G. Velarde, P. Colinet, \textit{Interfacial Phenomena and Convection}, Chapman and Hall/CRC, 2002. 
\bibitem{15}
V. G. Levich and V. S. Krylov, Annu. Rev. Fluid Mech. \textbf{1}, 293 (1969).
\bibitem{16}
J. L. Anderson, Annu. Rev. Fluid. Mech \textbf{21}, 69 (1989).
\bibitem{19}
Yu. S. Ryazantsev, Fluid Dynamics \textbf{20}, 491 (1985).
\bibitem{20}
Z. Izri, M.N. van der Linden, S. Michelin and O. Dauchot, Phys. Rev. Lett. \textbf{113}, 238302 (2014).
\bibitem{21}
T. Toyota, N. Maru, M.M. Hanczyc, T. Ikegami and T. Sugawara, J. Am. Chem. Soc. \textbf{121}, 5012 (2009).
\bibitem{22}
H. Kitahata, N. Yoshinaga, K. H. Nagai and Y. Sumino, Phys. Rev. E \textbf{84}, 015101 (2011).
\bibitem{23}
K. Nagai, Y. Sumino, H. Kitahata and K. Yoshikawa, Phys. Rev. E \textbf{71}, 065301 (2005).
\bibitem{26}
Y. Chen, Y. Nagamine and K. Yoshikawa, Phys. Rev. E \textbf{80}, 016303 (2009). 
\bibitem{29}
Y. Sumino, N. Magome, T. Hamada and K. Yoshikawa, PRL 94, 068301 (2005).
\bibitem{37}
M. Schmitt and H. Stark, Eur. Phys. J E \textbf{39}, 80 (2016).
\bibitem{38}
A. Zottl and H. Stark, J. Phys. Condens. Matter \textbf{28} 253001 (2016).
\bibitem{40}
M. Schmitt and H. Stark, Phys. of Fluids \textbf{28}, 012106 (2016).
\bibitem{41}
S. Yabunaka, T. Otha and N. Yoshinaga, J. Chem. Phys. \textbf{136}, 074904 (2012).
\bibitem{42}
S. Yabunaka and N. Yoshinaga, J. Fluid Mech. \textbf{806}, 205 (2016).
\bibitem{43}
N. Yoshinaga, Phys. Rev. E \textbf{89}, 012913 (2014).
\bibitem{44}
N. Yoshinaga, K. Nagai, Y. Sumino and H. Kitahata, Phys. Rev. E \textbf{86}, 016108 (2012).
\bibitem{45}
J. Bray, Adv. Phys. \textbf{43}, 357 (1994).
\bibitem{46}
A. Tiribocchi, N. Stella, G. Gonnella and A. Lamura, Phys. Rev. E \textbf{80}, 026701 (2009);
G. Gonnella, A. Lamura, A. Piscitelli, and A. Tiribocchi, Phys. Rev. E \textbf{82}, 046302 (2010).
\bibitem{47}
Z. Guo, C. Zheng, and B. Shi, Phys. Rev. E \textbf{65}, 046308 (2002).
\bibitem{48}
G. Kahler, F. Bonelli, G. Gonnella and A. Lamura, Phys. of Fluids \textbf{27}, 123307 (2015);
A. Coclite, G. Gonnella and A. Lamura, Phys. Rev. E \textbf{89}, 063303 (2014).
\bibitem{49}
A. Lamura, G. Gonnella and J.M. Yeomans, Europhys. Lett. \textbf{45} (3), 314 (1999); A. Lamura, G. Gonnella and J.M. Yeomans, Int. J. Mod. Phys. C \textbf{9} 8, 1469 (1998).
\bibitem{51}
T. Ishikawa, J. R. Soc. Interface (2009) \textbf{6}, 815-834;
A.A. Evans, T. Ishikawa, T. Yamaguchi and E. Lauga, Phys. Fluids \textbf{23}, 111702 (2011).
\bibitem{53}
M.T. Downton and H. Stark, J. Phys. Condens. Matter \textbf{21}, 204101 (2009).
\bibitem{54}
J. Elgeti, R.G. Winkler and G. Gompper, Rep. Prog. Phys. \textbf{78}, 056601 (2015).
\bibitem{52}
I. Goetze and G. Gompper, Phys. Rev. E \textbf{82}, 041921 (2010).
\end{thebibliography}
\end{document}